\documentclass[10pt,a4paper]{article}

\date{}
\begin{document}

\title{Local cosmology of the solar system\\
{\normalsize }}
\author{Ll. Bel\thanks{e-mail:  wtpbedil@lg.ehu.es}}

\maketitle

\begin{abstract}

A time-dependent model of space-time is used to describe  the gravitational field of the sun. This model is a spherically symmetric approximate solution of Einstein's equations in vacuum. Near the sun it approximates one of the models derived from the Schwarzschild solution, while at large distances it becomes a milne's-like zero space-time curvature model. Two local cosmology free parameters provide simple descriptions for the secular increasing of the astronomical unit, as well as the  "anomalous" radial acceleration  of the Pioneer probe. We make also a comment about the possibility of deriving MOND's phenomenology from General relativity.

\end{abstract}

\section{The Field equations}

Let us consider the following line-element, whose derivation will be summarized in the Appendix:

\begin{eqnarray}
\nonumber
&& \hspace{-2cm} ds^2:=-A(r,t)^2(-dt+atdr)^2 \\ [1ex]
\label{1.1}
&& \hspace{-0cm}+A(r,t)^{-2}\left(\frac{1+a\mu(r)t^2}{1+p^2r^2}
+(1+a\nu(r)t^2)\left(1-\frac{2m}{r}\right)r^2d\Omega^2\right)
\end{eqnarray}
where:

\begin{equation}
\label{1.2}
A(r,t)=\left(\left(1-\frac{2m}{r}\right)^{-1/2}+(1+2pt)^{1/2}-1\right)^{-1}
\end{equation}

\begin{eqnarray}
\label{1.3.1}
\mu(r)&=&\frac{3m}{r^2}-\frac{6m^2}{r^3}\left(1+\frac{2m}{r}\right)-p^2(m-r), \\
\label{1.3.2}
\nu(r)&=&\frac{1}{r}\left(1-\frac{4m}{r}\right)+\frac{4m^2}{r^3}\left(1+\frac{2m}{r}\right)-p^2(2m-r)
\end{eqnarray}
$m$ is the mass of the sun, $p$ and $a$ are two parameters. The units have been chosen so that the present values of $G$ and $c$ at $t=0$ are both 1.

If $a=0$ and  $p=0$ then the line-element above is that of Droste's model of Schwarzschild's solution. If $a=0$ and  $m=0$ then the line-element above is that of Milne's model of Minkowski's space-time. If $a=0$ but neither $m=0$ nor $p=0$ then the line-element is an approximate solution of Einstein's equations for vacuum that we discussed in \cite{Bel2010}. Finally, if $m=0$ but $p$ and $a$ are not zero the model it is akin to a modified Milne's model that we discussed in \cite{Bel2007}

Here we are interested in the case where none of the three parameters is zero. Let us consider a monomial such as:

\begin{equation}
\label{1.4}
a^{n_a}m^{n_m}p^{n_p}
\end{equation}
where the three exponents are zero or positive and such that $n_a<2,\ n_m<3$ and $n_p<3$. We shall say that the order of the monomial is $\epsilon=n_a+ n_m+n_p$, whatever be the values of the parameters involved.

A formal Taylor expansion of any non zero component of the Riemann tensor with the indices in the right position so that it has physical dimensions {\bf L${}^{-2}$} (or {\bf T${}^{-2}$}) has a leading term with a monomial of order 1, as for example:

\begin{equation}
\label{1.4.1}
R^1_{\ 414}=-2\frac{m}{r^3}
\end{equation}

On the other hand, formal Taylor expansions of the  non identically zero components of the Einstein tensor of the line-element (\ref{1.1}) lead to the following results, correct to order 3:

\begin{eqnarray}
\label{1.5.1}
S^0_1&=&-\frac{4mpa}{r^3}t^2 \\
\label{1.5.2}
S^0_0&=&\frac{4mp^2}{r}+a\left(\frac{16p^2}{r}t^2+\frac{2mp}{r^4}t^3\right)
\end{eqnarray}

\begin{eqnarray}
\nonumber
&& \hspace{-2cm} S^1_1=-\frac{2mp^2}{r}-\frac{2m^2p}{r^4}t
\\[1ex]
\label{1.5.3}
&& \hspace{-1cm}+a\left(
\left(
\frac{16p}{r}-\frac{52mp}{r^2}\right)t
+\left(
-\frac{8p^2}{r}-\frac{2m}{r^4}+\frac{10m^2}{r^5}\right)t^2+\frac{4mp}{r^4}t^3\right)
\end{eqnarray}

\begin{eqnarray}
\nonumber
&& \hspace{-2cm} S^2_2=-\frac{mp^2}{r}+\frac{2m^2pt}{r^4} \\
\\[1ex]
\label{1.5.4}
&& \hspace{-1cm} +a\left(
\left(
\frac{8p}{r}+\frac{4mp}{r^2}\right)t
+\left(
-\frac{4p^2}{r}+\frac{2m}{r^4}-\frac{16m^2}{r^5}\right)t^2-\frac{4mp}{r^4}t^3\right)
\end{eqnarray}

\section{The equations of motion}

Let us consider a test-particle, i.e. here a body with negligible mass with respect to the mass of the sun, and moving with a non relativistic velocity. The force per unit mass in the radial direction acting on this particle in the gravitational field of the sun, if it is described by the lie-element (\ref{1.1}), is:

\begin{equation}
\label{1.6}
f(r,t)=-\Gamma^1_{00}(r,t)A(r,t)^{-2}
\end{equation}
where $A(r,t)$ was defined in (\ref{1.2}) and $\Gamma^1_{00}(r,t)$ is the corresponding Christoffel symbol:

\begin{eqnarray}
\nonumber
&& \hspace{-2cm} \Gamma^1_{00}(r,t)=\frac{m}{r^2}-\frac{2m^2}{r^3}+mp^2-\left(\frac{5pm}{r^2}-\frac{15pm^2}{r^3}\right)t+\frac{35mp^2t^2}{2r^2}
\\[1ex]
\label{1.6.1}
&& \hspace{-1cm} +a\left(1-\frac{4m}{r}+\frac{4m^2}{r^2}+p^2r^2-\left(5-\frac{25pm}{r}\right)t-\left(\frac{3m^2}{r^4}-18p^2\right)t^2\right)
\end{eqnarray}

Using the formal expansion of $f(r,t)$ to order 2 yields then the following  equations of motion:

\begin{equation}
\label{1.7.1}
\ddot{r}-r\dot{\varphi}^2=-\frac{m}{r^2}(1-3pt)-a(1-\frac{2m}{r} -3pt)
\end{equation}
and:

\begin{equation}
\label{1.7.2}
2\dot{r}\dot{\varphi}+r\ddot{\varphi}=0 \ \ \hbox{or} \ \ \ r^2\dot{\varphi}=L
\end{equation}
$L$ being the constant angular momentum per unit mass.

Let us assume first that the test-particle is moving along a radial direction. Then we have, using (\ref{1.7.1}) with a further approximation to order $\epsilon=1$

\begin{equation}
\label{1.8}
\ddot{r}=-\frac{m}{r^2}-a
\end{equation}
This describes the dynamics of the Pioneer probe if, \cite{Anderson}:

\begin{equation}
\label{1.9}
a=(8.74 \pm 1.33)\times 10^{-10}\, \hbox{m/s}{}^2
\end{equation}

If we assume now that the test-particle is very slowly spiraling out from a circular orbit ($r\ddot r\ll 1,\dot r<1$) we obtain using (\ref{1.7.2}):

\begin{equation}
\label{1.10}
L^2=(1-3pt)r^3\left(\frac{m}{r^2}-a\right)+2ma
\end{equation}
Neglecting now $a$ compared to $m/r^2$ and differentiating this equation we get again the result of \cite{Bel2010}:

\begin{equation}
\label{1.11}
\dot r=\frac{3pr}{1-3pt}
\end{equation}
that with:

\begin{equation}
\label{1.12}
p=5\times 10^{-21}\,\hbox{s}^{-1}
\end{equation}
would describe an increasing of the AU of the order of 7\, m/cy (\cite{Claus}, \cite{Iorio}).

\section{Validation of the approximation}

Let us consider a domain of space-time $\cal D$ defined by these two intervals:

\begin{equation}
\label{1.13}
r_{min}<r<r_{max},\ \ 0<t<t_{max}
\end{equation}
where $r_{min}=AU$, $r_{max}=100\times AU$ and $t_{max}=cy$. Then,
with the values above for $a$ and $p$ we have:

\begin{eqnarray}
\label{1.14}
\frac{m}{r_{min}}\approx 10^{-8}, \ pr_{max}\approx 10^{-16}, \ ar_{max}\approx 10^{-13} \\
pt_{max}\approx 10^{-11},  \ at_{max}\approx 10^{-8}
\end{eqnarray}
These are all very small numbers but the number that is really conditioning the quality of the approximation is:

\begin{equation}
\label{1.15}
Q=at_{max}\frac{t_{max}}{r_{min}}\approx 10^{-3}
\end{equation}
that depends on the size and the shape of the  domain $\cal D$ through the dimensionless parameter $t_{max}/r_{min}$ and that in our example is of the order of $10^7$. $Q$ measures the deviation of the line-element (\ref{1.1}) from the corresponding one with $a=0$. It measures
also the dimensionless ratio of the greatest value of the components of the Einstein tensor and the greatest value of the leading terms of the components of the Riemann tensor\,\footnote{In \cite{Bel1987} we gave a general definition of the Quality factor of an approximate model of space-time}.

\section{Comment}

It is customary to say that the physics of galaxies is essentially Newtonian physics: The velocities involved are non relativistic and the gravitational fields are weak. We can add to these reasonable considerations some more facts. Let us consider for instance one of them. NGC 4472 is a huge spherical galaxy which has a proper mass of the order of $10^{12}$ solar masses and a radius of about 25 kpc. Despite these data the compactness parameter of this galaxy, $\lambda=3.8\,10^{-6}$, is surprisingly close to the compactness parameter of the sun, $\lambda=4.3\,10^{-6}$:

Can we conclude from the preceding remarks that in our universe the size of isolated systems do not matter if the velocities are non relativistic, the gravitational fields are weak and have comparable compactness parameters? The answer to this question is: certainly not if we use for all of them the same model that we have used here to describe the solar system.

The following example tells us why. As far as 1oo au from the sun the strength of the gravitational field is the small quantity $g=6\,10^{-6}$ m/s${}^2$ while at 25 kpc of the center of NGC 4472 the strength of the gravitational field is the much smaller quantity $g=2.2\,10^{-10}$ m/s${}^2$. In the first case $g$ remains larger than the parameter $a$ while in the second case $g$ is coincidentally almost equal to $a$.

Let us examine how gravitationally bound objects would describe circular orbits around  galaxies like NGC 4472  or similar. If $r_s$ is the radius of the galaxy and $v_s$ is the speed at this distance from the center, then the sped $v$ of an object orbiting it at a distance $r_s+\delta r$ would be, to first order in $\delta r$:

\begin{equation}
\label{A.5}
v=v_s\left(1-\eta\frac{\delta r}{2r_s}\right) \quad \hbox{with}\quad  \eta=1-\frac{2ar_s}{v_s^2}
\end{equation}
Small values of $\eta$ would mean approximate flatness of the rotation curves suggesting that MOND's {\it ad hoc} proposition, \cite{Milgrom1983}, could after all be derived from a suitable approximated solution of Einstein's field equations in vacuum.

\section*{Appendix}

Let us consider the line-element (\ref{1.1}) without specifying the functions $\mu(r)$ and $\nu(r)$, the function $A(r,t)$ being still be given by (\ref{1.2}). The Cauchy data of this line-element on a hyper-surface $t=0$ depends only on:

\begin{equation}
\label{A.1}
A(r,0)=\left(1-\frac{2m}{r}\right)^{1/2}, \ \partial_t A(r,0)=-p\left(1-\frac{2m}{r}\right)
\end{equation}

From Lichnerowicz's \cite{Lichne} analysis of the Cauchy problem  we know that $S^0_1$ and $S^0_0$ for $t=0$ will not depend on $\mu(r)$ or $\nu(r)$. Actually we have neglecting terms of order 4 and higher, as we have been doing:

\begin{equation}
\label{A.2}
S^0_1(r,0)=0, \ \ S^0_0(r,0)=\frac{4mp^2}{r}
\end{equation}
The terms of order 3 are maintained to visualize the first neglected terms but the approximation, that we work with, considers them to be negligible. Therefore we say that the problem of the inial conditions is already solved.

On the other hand, to discuss the evolution of the line-element in a neighborhood of $t=0$ requires, as a first step, to consider the two components of the Ricci tensor:

\begin{eqnarray}
\nonumber
&& \hspace{-2cm} R_{11}(r,0)=-\frac{a}{2} \left(1+4m -p^2*r^2+\frac{12m^2}{r^2}\right)\mu(r) \\[1ex]
\label{A.3.1}
&& \hspace{+3cm}+\frac{3m}{r^2}+\frac{6m^2}{r^3}+p^2r -\frac{2p^2m}{r}\\[1ex]
\label{A.3.2}
&& \hspace{-2cm} R_{22}(r,0)=-\frac{a}{2} \left((r^2+mr+2m^2)\nu(r)-2m+p^2r^3\right)-p^2mr
\end{eqnarray}
Solving the equations:

\begin{equation}
\label{A.4}
R_{11}(r,0)=-\frac{2p^2m}{r}, \ \ R_{11}(r,0)=-p^2mr
\end{equation}
This eliminates the terms proportional to $a$ and yields the expressions (\ref{1.3.1}) and (\ref{1.3.2}) for $\mu(r)$ and $\nu(r)$ that we used in the first section.

Thus, the space-time model that we have been using in the main body of this paper can be considered as an educated guess derived from previous work in \cite{Bel2010} and from an analysis of a first step  of a Cauchy problem.

\end{document}